\pdfoutput=1
\documentclass[alpha-refs]{wiley-article}

\usepackage{color}
\usepackage{ulem}

\usepackage{graphicx}
\usepackage{amsmath}
\usepackage{bm}
\usepackage{xcolor}
\usepackage{comment}
\usepackage{booktabs}
\usepackage{multirow}
\usepackage{siunitx}

\usepackage{hyperref}
\hypersetup{colorlinks=true, citecolor=blue, linkcolor=blue}

\title{Improving the representation of the atmospheric boundary layer by direct assimilation of ground-based microwave radiometer observations}

\author[1\authfn{1}]{Jasmin Vural}
\author[2\authfn{1}]{Claire Merker}
\author[3,4]{Moritz L\"offler}
\author[2]{Daniel Leuenberger}
\author[1]{Christoph Schraff}
\author[1]{Olaf Stiller}
\author[1]{Annika Schomburg}
\author[5]{Christine Knist}
\author[6]{Alexander Haefele}
\author[6]{Maxime Hervo}

\contrib[\authfn{1}]{Equally contributing authors.}

\affil[1]{Research and development, Deutscher Wetterdienst, Offenbach, Germany}
\affil[2]{Numerical Prediction, MeteoSwiss, Z\"urich, Switzerland}
\affil[3]{Technical Infrastructure and operations, Deutscher Wetterdienst, Potsdam, Germany}
\affil[4]{Institute for Meteorology and Geophysics, University of Cologne, Cologne, Germany}
\affil[5]{Meteorological Observatory Lindenberg - Richard-A{\ss}mann-Observatory (MOL-RAO), Deutscher Wetterdienst, Tauche, Germany}
\affil[6]{Atmospheric Measurements, MeteoSwiss, Payerne, Switzerland}

\corraddress{Annika Schomburg, Deutscher Wetterdienst, Offenbach 63067, Germany.}
\corremail{annika.schomburg@dwd.de}

\fundinginfo{Jasmin Vural, German Research Foundation (DFG), Grant Number: 399851006}

\runningauthor{accepted for publication in QJRMS 2023-12-07}

\begin{document}
\maketitle
\begin{abstract}
In a joint effort, MeteoSwiss and Deutscher Wetterdienst (DWD) address the need for improving the initial state of the atmospheric boundary layer (ABL) by exploiting ground-based profiling observations that aim to fill the existing observational gap in the ABL. We implemented the brightness temperature observations from ground-based microwave radiometers (MWRs) into our data assimilation systems using a local ensemble transform Kalman filter (LETKF) with RTTOV-gb (Radiative Transfer for TOVS, ground-based) as a forward operator. We were able to obtain a positive impact on the brightness temperature first guess and analysis as well as a slight impact on the ABL humidity using two MWRs at MeteoSwiss. These results led to a subsequent operational implementation of the observing system at MeteoSwiss. Furthermore, we performed an extensive set of assimilation experiments at DWD to further investigate various aspects such as the vertical localisation of selected single channels. We obtained a positive impact on the \SI{6}{\hour}-forecast of ABL temperature and humidity by assimilating two channels employing a dynamical localisation based on the sensitivity functions of RTTOV-gb but also with a static localisation in a single-channel setup. Our experiments indicate the importance of the vertical localisation when using more than one channel, although reliable improvements are challenging to obtain without a larger number of observations for both assimilation and verification.


\end{abstract}

\section{Introduction}

The atmospheric boundary layer (ABL) can be observed using different in-situ and remote-sensing instruments. Apart from information on clouds, it is difficult to extract information on the ABL from satellite observations, especially over land. The observations typically used, including radiosonde and aircraft data, however, provide only a very sparse spatial or temporal resolution, especially for vertical profiles. This severely affects the forecast quality of numerical weather prediction (NWP) and additional observations seem critical for improving the NWP, particularly in ABL-dependent weather conditions such as weakly forced convection, low-stratus or fog as these are highly dependent on the initial conditions and their uncertainties \citep{2001clahop,2005bercar}.

Currently, several European initiatives like the E-PROFILE project by EUMETNET \citep{2021rufhae}, ACTRIS \footnote{\url{https://www.actris.eu/}}, or the COST action PROBE \citep{2020cimhae} aim at filling the observational gap in the ABL to address the needs of weather services, science, and industry for an improved knowledge of the state of the ABL. The effort is focused on the extension of ground-based networks for ABL profiling, an improved quality of the provided data, and a better exploitation of the data in applications, for example in NWP, where additional profiler observations can be beneficial to the initial conditions \citep[e.g.][]{2019illcim,2020leuhae}.

In accordance with these activities, the two national meteorological centres (NMC) of Germany (Deutscher Wetterdienst, DWD) and Switzerland (MeteoSwiss) investigated the benefit of assimilating different novel ground-based remote-sensing profiling instruments, especially Doppler wind lidars, Raman lidars, and microwave radiometers (MWRs).
Among these remote sensing systems, the MWRs are among the most advanced in our evaluation in terms of technical operation and quality, and we present here our advancements and results on the assimilation of MWR observations into our regional NWP systems.

MWRs are passive remote sensing instruments, which can in principal be spaceborne as well. Here, we refer to the ground-based instruments only, which measure downwelling radiation. The radiation spectra contain information on the air temperature and water vapour content in the atmospheric column above the instrument up to a height of \SI{\approx 3}{\kilo \metre} \citep{2005roscre, 2006cimhew}, and therefore have the potential to improve the representation of the ABL in NWP through data assimilation \citep{2017marcim,2019illcim,2020chiwan}. 
At DWD, currently one MWR is operated within a project called ``Pilotstation'', where also other ground-based profilers, such as different lidars (Doppler wind lidar, water vapour broadband differential absorption lidar, and in the future, a compact Raman lidar), and a cloud radar, are operated and evaluated for their technical readiness for operational data assimilation. On its part, MeteoSwiss routinely operates three MWR as part of an integrated analysis and forecasting system for emergency response, e.g., to nuclear incidents. 

Vertically resolved temperature and humidity profiles can be retrieved from the observed brightness temperatures with inverse (non-scattering) radiative transfer models under certain assumptions \citep{2007turclo,2015kos}.
Retrieved MWR profiles of temperature and humidity were already assimilated in an NWP model by \citet{2016caucim} who obtained a small positive impact on the precipitation forecast up to \SI{18}{\hour} in a rapid update cycle system. \citet{2020marcim} demonstrated a positive impact of the retrieved profiles on the initial state of the NWP model in foggy conditions within a 1D-VAR system. In contrast, \citet{2023linsun} reported difficulties with the assimilation of retrievals due to difficulties with the bias correction.

In this article, we present a different approach of using MWR observations, namely by assimilating the measured brightness temperatures directly in order to circumvent the assumptions inherent to retrieved profiles. These assumptions rely on auxiliary data (model or observations) and/or statistically derived coefficients that cannot address flow-dependent weather situations \cite[cf.][]{2012lohmai}.

The work presented here was done in collaboration between DWD and MeteoSwiss.
The paper is structured as follows: In Sect.~\ref{sec:obs}, we present the MWR instruments and the observational pre-processing, Sect.~\ref{sec:setup} describes the common data assimilation system, the integration of MWR observations into the system employing the forward operator, and an assessment of the observation quality by comparison with the model background.
Section~\ref{sec:mch_setup} presents the operational assimilation of MWR observations at MeteoSwiss and Sect.~\ref{sec:exps_dwd} the continued investigations to enhance and better understand the impact on our forecast systems at DWD.

\section{Microwave radiometer observations}\label{sec:obs} 

MeteoSwiss and DWD use instruments at different sites.
DWD operates one MWR, which is located at the Meteorological Observatory Lindenberg - Richard-A{\ss}mann-Observatory (MOL-RAO), Germany, where a wealth of other atmospheric measurements is provided. Most important for our purposes are the radiosonde ascents every six hours. These observations serve as a reference for the forecast verification.

MeteoSwiss operates three MWR instruments in Payerne, Grenchen, and Schaffhausen (data from the latter is not used here because the device is of older generation). The observational site in Payerne also features radiosounding ascents and, among other instruments, a Raman lidar \citep{2013dinsim,2013brophi}. Those profiling instruments are used as reference for the assessment of the NWP quality.

At all sites, the installed MWR is the HATPRO-G5 by Radiometer Physics GmbH (RPG; \citealt{2005roscre}). It observes the atmospheric radiance as brightness temperatures ($T_\mathrm{b}$).
The $T_\mathrm{b}$ are observed on 14 channels in the K- and V-band (Table~\ref{tab:mwr_channels}). The seven K-band channels (\SIrange{22.2}{31.4}{\giga \Hz}; numbered as 1 to 7 in the following) are located on the humidity-sensitive water vapour absorption feature, whereas the seven V-band channels (\SIrange{51.3}{58.0}{\giga \Hz}; numbered 8 to 14) lie on the temperature-sensitive oxygen absorption feature. All channels below \SI{\approx 56}{\giga \Hz} are sensitive to cloud liquid water. 

Ground-based MWRs observe $T_\mathrm{b}$ during cloudy and clear-sky conditions and are sensitive up to a height of \SI{\approx 3}{\kilo \metre} \citep{2006cimhew}. Being passively sensed, the observations can be vertically located but are not vertically resolved in a strict sense. 
The degrees of freedom of MWR retrievals are limited by the number of channels and the overlapping information content between the channels \citep{2021turloh}. 
The HATPRO devices are able to scan the ABL with different elevation angles. The elevation scans can add more degrees of freedom to the system, however, we considered only zenith observations for the studies presented here. A comparison between measured brightness temperatures and model values performed for the slanted scans showed larger discrepancies than for the zenith scan. This can be partially attributed to the fact that the forward operator RTTOV-gb (see Sect.~\ref{sec:rttovgb}) assumes a homogeneous atmosphere for the slanted scans. Removing this simplification from the forward model is expected to be beneficial especially for the temperature profiles and will be subject of future work.  Additionally, the cloud detection for the elevation scans is more challenging, and, eventually, the error cross-correlations between brightness temperatures of different elevation angles cannot be treated yet by the assimilation system.

The MWR observations are provided with a temporal resolution down to one second. To smooth out short-term fluctuations that cannot be resolved by the model data, we average the observations over one minute at DWD, and over five minutes at MeteoSwiss. The values differ due to different technical standards at the NMCs. A longer averaging interval might be slightly more representative for the model grid, whereas the shorter interval will generally be less contaminated with cloudy observations.

The presence of clouds decreases the representativeness between model and observation in some channels. Observations in cloudy conditions must therefore be treated separately (see Sect.~\ref{sec:scexp}). 
Cloudy conditions at the sites are detected employing different techniques. At MeteoSwiss, the MWR sites are equipped with ceilometers used for cloud detection. At DWD, two different algorithms are used. One is based on the brightness temperature in the infrared (\SI{10.5}{\micro \metre}; $T_\mathrm{b,IR}$) measured by a infrared radiometer mounted on the MWR. Clouds are assumed for $T_\mathrm{b,IR}>$ \SI{243.14}{\kelvin} as in \citet{2015mardab}, motivated by the statistical prevalence of cloud liquid water above \SI{243.14}{\kelvin} \citep{2003hogill,2014kohgor}. In addition, the standard deviation $\sigma$ of the \SI{31.4}{\giga \Hz} channel is evaluated in a time window of \SI{\pm 3}{\minute} around the timestamp of interest. Clouds are assumed when $\sigma(T_\mathrm{b},\SI{31.4}{\giga \Hz}) >$ \SI{0.25}{\kelvin}. We assume clear-sky when both methods detect no liquid cloud.

During and partially also after rain, the observed $T_\mathrm{B}$ are biased. To mitigate this effect, the MWRs are equipped with a hydrophobic radome which must be exchanged every few months. Rainfall is detected with an acoustic disdrometer \citep{2011salelo} and observations during rain events are discarded \citep{1996she}. If the hydrophobic property of the radome has degraded, remaining water will disturb observations up to several minutes after the rainfall event has stopped. At DWD, these cases are detected with a spectral consistency check and discarded.

The MWR HATPRO-G5 must be calibrated with liquid nitrogen every six months. Calibration is required regularly to account for possible drifts of the receiver characteristics and subsequently biased $T_\mathrm{b}$ \citep{2016kuctur}. 

\begin{table*}[htb]
\setlength{\tabcolsep}{2.7pt}
\caption{Band names, channel numbers, and frequencies (in \SI{}{\giga\hertz}) of the HATPRO-G5 MWR.} \label{tab:mwr_channels}
    \centering
    \begin{tabular}{l|ccccccc|ccccccc}
    \toprule
    Band     & \multicolumn{7}{c|}{K} & \multicolumn{7}{c}{V} \\
    \midrule
    Channel & 1 & 2 & 3 & 4 &  5 & 6 & 7 & 8 & 9 & 10 & 11 & 12 & 13 & 14 \\
    Frequency & 22.24 & 23.04 & 23.84 & 25.44 & 26.24 & 27.84 & 31.40 & 51.26 & 52.28 & 53.86 & 54.94 & 56.66 & 57.30 & 58.00 \\
    \bottomrule
    \end{tabular}
\end{table*}

\section{The data assimilation systems} \label{sec:setup}

\begin{figure}[htb]
    \includegraphics[width=0.5\textwidth]{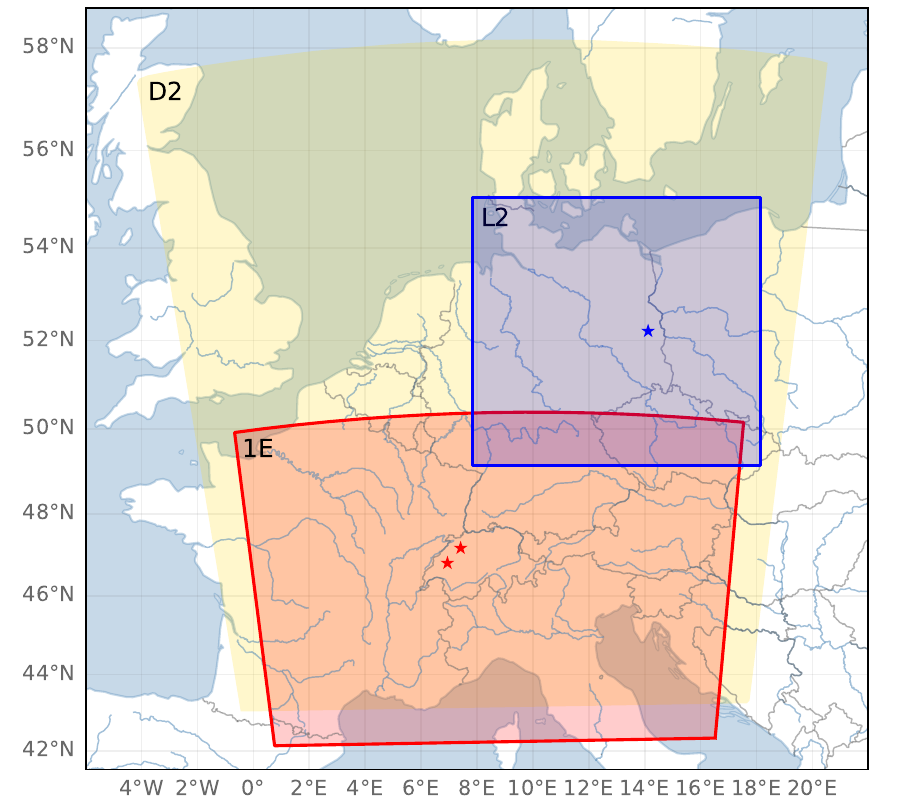}\\
    \caption{Illustration of the employed domains. The standard ICON-D2 domain (D2; yellow), the smaller ICON domain (L2; blue) around the MOL-RAO (blue star), and the COSMO-1E domain (1E; red) with the MWR stations Payerne and Grenchen (red stars).}\label{fig:dom}
\end{figure}

\subsection{Method and models} \label{sec:model_setup}

The kilometre-scale ensemble data assimilation system \citep[KENDA;][]{2016schrei,2016bicsim} is based on the local ensemble transform Kalman filter \citep{2007hunkos}. It is a convective-scale, ensemble data assimilation system that can be coupled to two different NWP models, the COSMO \citep{2021dombal,2021domfor} and the ICON model \citep{2015zanrei}. 
The NWP model provides the first-guess (FG; also called background) of the initial state, a forecast with one hour lead time, that is updated hourly to an analysis by the LETKF using observations. In our setups, KENDA runs with 40 members plus one deterministic run. In all FG runs, radar-derived surface precipitation rates are assimilated by latent heat nudging. The horizontal localisation radius of an observation is adapted dynamically between \SI{50}{\kilo \metre} and \SI{100}{\kilo \metre} dependent on the density of other observations in the vicinity.

The NWP models used in operations at DWD and MeteoSwiss currently differ. DWD already performed the migration to the recently developed ICON model whereas MeteoSwiss is in the midst of this process and still runs COSMO routinely. Therefore, the setup of the data assimilation systems used for the presented experiments differ between the two NMCs (Table~\ref{tab:das}). The data assimilation system and the coupling to the forward operator RTTOV-gb (see Sect.~\ref{sec:rttovgb}), however, are identical. 

Due to the different regions of interest, we ran our experiments on different domains (Fig.~\ref{fig:dom}). MeteoSwiss uses the full COSMO-1E domain, with lateral boundary conditions taken from the European Centre for Medium-Range Weather Forecast (ECMWF) global deterministic (Integrated Forecasting System at high resolution; IFS HRES) and ensemble (IFS-ENS) models. At DWD, a smaller domain has been created as computations on the full ICON-D2 domain are quite expensive in terms of computational resources. This subdomain (L2) is centred a bit south west of Berlin and has an extent of approximately \SI{500}{\kilo \metre} x \SI{500}{\kilo \metre}. As for the standard ICON-D2 domain, the boundary conditions were generated with the ICON-EU model, which has a horizontal mesh-width of \SI{6.5}{\kilo \metre}. 

\begin{table*}[htb]
\caption{Setup of observations, assimilation system, and forward operator at both NMCs. The parameter $\alpha_\mathrm{bias}$ represents a weight in the dynamical bias correction and will be explained in Sect.~\ref{sec:omb}.} \label{tab:das}
    \centering 
    \begin{tabular}{l|ll}
    \toprule
                            & MeteoSwiss        & DWD                     \\ \midrule
      MWR stations          & Payerne, Grenchen & MOL-RAO                              \\
      station elevation     & \SI{490}{\metre}, \SI{428}{\metre} & \SI{127}{\metre}    \\
      MWR assimilation frequency & hourly       & hourly                               \\
      NWP model             & COSMO-1E          & ICON-D2                              \\
      grid spacing& \SI{\approx 1.1}{\kilo \metre}  & \SI{\approx 2.1}{\kilo \metre} \\
      $\alpha_\mathrm{bias}$& 0.0015            & 0.004                                \\
      vertical localisation & see Sect.~\ref{sec:mch_setup} & see Sect.~\ref{sec:vloc} \\
      clouds                & see Sect.~\ref{sec:mch_setup} & see Sect.~\ref{sec:vloc} \\
    \bottomrule
    \end{tabular}
\end{table*}

\subsection{The forward operator RTTOV-gb} \label{sec:rttovgb}

To assimilate the MWR brightness temperatures, the model variables are transformed into observation space by applying the radiative transfer model RTTOV-gb \citep{2016decim} as forward operator to the first guess provided by the NWP model. The model profiles (temperature, specific humidity, cloud liquid water, pressure), and model screen-level variables (\SI{2}{\metre} temperature, surface pressure) are used to compute model brightness temperatures $T_\mathrm{b}$ that can be compared directly to the MWR observations. Clear- and all-sky computations are both possible due to the incorporation of the absorption of liquid cloud water in RTTOV-gb.

As described in \citet{2016decim}, the input model levels are interpolated to the 101 fixed layers of RTTOV-gb that have an enhanced vertical resolution in the lower atmosphere. The relationship between optical depth and the input state vector profile is described by a linear combination of predictors, the coefficients were derived by linear regression. In addition, RTTOV-gb provides the sensitivities on the model variables, i.e., the Jacobians \citep{2016decim}, which can be used for determining the height assignment necessary for the vertical localisation of the single channels (see Sect.~\ref{sec:vloc}).

We developed a wrapper for the forward-model computation, allowing for flexible adjustments of, e.g., channel selection, bias correction, observation errors, cloud treatment, vertical localisation, and elevation angles. We implemented this so-called model equivalent calculator into the KENDA data assimilation system at DWD and MeteoSwiss with slightly different settings each (Table~\ref{tab:das}).

\subsection{O-B statistics} \label{sec:omb}

In preparation for the assimilation of the MWR observations, we compare time series of both observations and background $T_\mathrm{b}$ by evaluating so-called observation minus background (O-B, also called first guess departures) statistics. This is crucial to assess the deviations between model and observations, and thus, the suitability of the observations for data assimilation, and is also required here to estimate parameters needed for the assimilation itself. In addition, O-B statistics enable us to assess the general behaviour of the different channels, also with regard to cloudy and clear-sky conditions. 

We employed the O-B standard deviation (STD) for the derivation of a first estimate for the observation errors of each channel. In a data assimilation system, the observation error should not be considered as the measurement accuracy only, but rather as a measure that also includes the representativeness error between observation and model, and errors in the observation operator. We analysed clear-sky and cloudy conditions separately and found that the O-B STD is strongly increased during cloudy conditions for all channels except channel 10 to 14. This information can be used to assign different observation errors to the different conditions.

First estimates of the systematic errors for each channel are derived by computing O-B temporally averaged over a certain period. For the DWD experiments in this study, we averaged over the investigated period. 
 To include new devices in the operational setup of MCH, the data available between the installation of the device and the integration in the data assimilation system, typically one to two months, are used for this purpose. These estimates represent the biases of the observations with respect to the model and are used as initial values for the adaptive bias correction, which is applied to the observations prior to assimilation at each analysis step. A bias correction of observations aiming to adapt the climatology of the observations to the model's climatology is common practice in data assimilation, and in doing so meet the bias-free assumption of the LETKF (e.g., \citealt{1998deeda}). The value for bias correction $\langle d^o_b\rangle_t$ is adapted for each analysis at time $t$ during the assimilation cycle by adding the weighted current O-B value $d^o_b(t)$ to the bias of the last time step $\langle d^o_b\rangle_{t-1}$:
\begin{equation}
\langle d^o_b\rangle_t = \alpha_\mathrm{bias}\, d^o_b(t) + (1 - \alpha_\mathrm{bias}) \langle d^o_b\rangle_{t-1} \;. \label{eq:bc}
\end{equation}
The weighting factor $\alpha_\mathrm{bias}$, basically a tuning factor determining the speed at which the bias correction factor adapts to drifts or jumps in the values of the observations, is set as indicated in Table~\ref{tab:das}.

\begin{figure*}[h!]
\includegraphics[width=\textwidth]{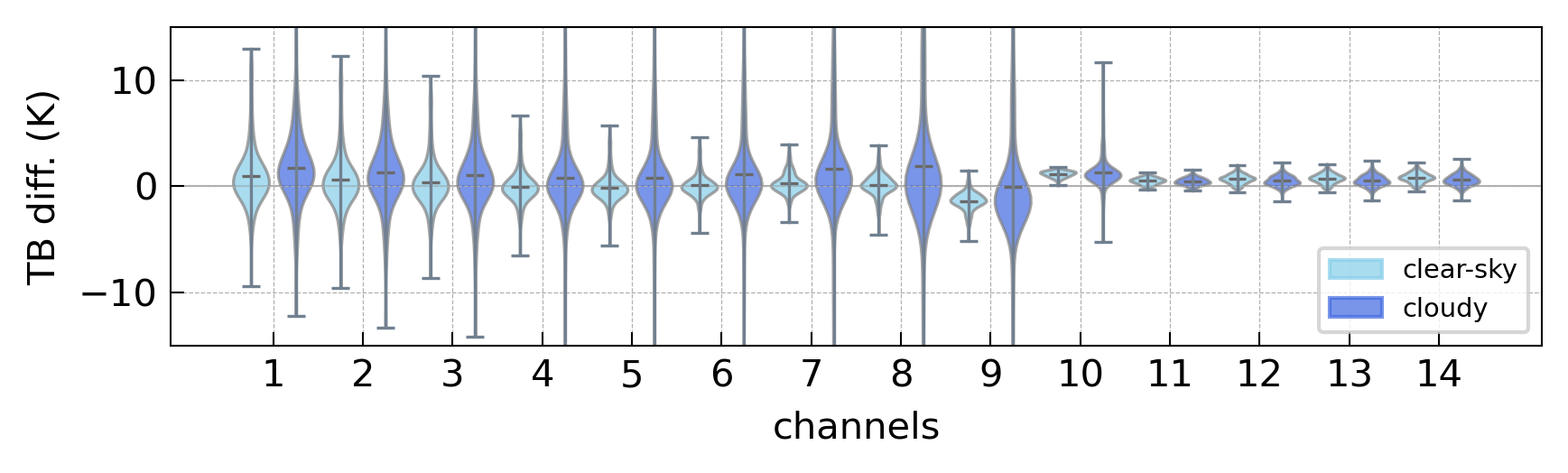}
    \caption{Observation minus background (O-B) statistics for the brightness temperature of the zenith scan of the MWR in Payerne. The violin plots show the distributions of the brightness temperature departures for each of the 14 channels, in clear-sky (bright blue) and cloudy (dark blue) conditions. The mean, maximum, and minimum values of each channel are indicated by a horizontal bar. The range of the y-axis is adapted to the clear-sky data, outlier values in cloudy cases can be much larger.} \label{fig:box_payerne}
\end{figure*}

The different settings of our systems do not allow for a direct comparison of the O-B statistics of all the employed MWRs, mainly due to the different models, model domains, and investigated time periods. The general behaviour of the $T_b$ for the different channels, especially with regard to clear- and all-sky is very similar though. We would like to refer here to \citet{2017decim} who showed that different MWR across Europe show a reasonable consistency between their respective O-B statistics.

Exemplary, we discuss the O-B statistics for one MWR (Payerne) gathered for the time period 15. February 2022 to 31. July 2022. The model equivalents are computed from the deterministic first guess run of the operational MeteoSwiss KENDA cycle (see Sect.~\ref{sec:model_setup} and \ref{sec:mch_setup} for more details). The violin plots in Fig.~\ref{fig:box_payerne} show the distributions of the O-B values for the 14 channels of the MWR, for clear-sky and cloudy (i.e. clouds in the observations or/and in the model) conditions separately. In general, most of the O-B values follow a normal distribution, but the distributions feature asymmetrical tails that affect the computed mean values. The K-band channels have a larger spread in the deviations between observations and model values than the V-band channels, in general. Especially channels 10 to 14, which are mainly sensitive to the atmospheric temperature, show a very good agreement, even though a small bias is apparent. This bias can be corrected using the adaptive bias correction described in Eq.~\ref{eq:bc}. 
Larger spectroscopic uncertainties present in the absorption models used in RTTOV-gb affect mainly frequencies between \SIrange{45}{54}{\giga\hertz} \citep{2018cimros} and, thus, larger deviations between observations and background for channel 8 an 9 can be attributed to these uncertainties in the forward model.
In addition, the O-B standard deviation of the channels \SIrange{1}{10}{} is also impacted by the presence of clouds. This is expected since humidity and especially clouds are strongly variable in time and space and therefore more prone to model errors than temperature. 

\section{Operational assimilation of MWR observations at MeteoSwiss} \label{sec:mch_setup} 

As described in Sect.~\ref{sec:model_setup}, MeteoSwiss runs an hourly data assimilation cycle using KENDA with the NWP model COSMO. The model domain is centered around the Alpine Arch (Fig.~\ref{fig:dom}) and the model grid spacing is approximately \SI{1.1}{\kilo \metre} (Table~\ref{tab:das}) for the data assimilation cycle (first guess and analysis) and the forecasts. The following observations are assimilated: wind, humidity and temperature profiles from radiosoundings (mainly at \SI{00}{UTC} and \SI{12}{UTC}); Doppler wind radar data from Swiss stations Payerne, Grenchen, and Schaffhausen, as well as a few stations outside of Switzerland, at half-hourly intervals; surface pressure, \SI{2}{\metre} temperature and \SI{2}{\metre} relative humidity from surface stations of the SwissMetNet \citep{2003fre} and partner stations in Germany, France, Austria and Italy, also at half-hourly intervals; aircraft reports of temperature and wind; surface precipitation estimates from the five Swiss radar stations at five-minute intervals using latent heat nudging \citep{1997jonmac,2007leuros,2008stekli}. 

A first technical development phase aimed at integrating the forward-operator RTTOV-gb into the workflow and including MWR observations into the KENDA data assimilation systems. After an assessment of the quality of MWR observations and of the deviations between measured and simulated $T_\mathrm{b}$, and the subsequent evaluation of assimilation experiments, MeteoSwiss introduced the MWR assimilation into the operational NWP setup in the summer of 2022. The setup aligns with the one described in Sect.~\ref{sec:setup} and Table~\ref{tab:das}. Observation errors and initial values for the adaptive bias correction were determined based on pre-operational, clear-sky O-B statistics (Table~\ref{tab:mch_mwr_settings}). To simplify the operational setup, and because the observation errors derived from the O-B statistics for both employed MWRs are similar in the evaluated data sample, the same observation errors are specified for both MWR devices. Additionally, a pragmatic approach is applied to address larger deviations between observations and the model under cloudy conditions (i.e. clouds in the observations or in the model). After some tuning, we decided to multiply the clear-sky observation error values by two for those cases.

Channels 2, 3, 7 and 10 to 14 are also assimilated in cloudy conditions. This maximises the availability of MWR observations for data assimilation and did not show to be detrimental in performed experiments. The current setup does not include vertical localisation for the MWR observations. The height range in which the MWR observations impact the model analysis is therefore determined by the LETKF using the ensemble covariance matrices (refer to Sect.~\ref{sec:vloc} for more details).

Moreover, it should be mentioned that channels 5, 6, 8 and 9 are excluded from assimilation. For these channels, O-B statistics from the MWR in Grenchen showed larger discrepancies in the analysed period (not shown here), leading to the decision to remove them from the operational assimilation setup. The MWR in Grenchen has since been replaced and its quality has improved significantly. Nevertheless, the setup remains unchanged since we assume that the assimilated channels already contain all the information that can be extracted from the data by the data assimilation scheme, especially since we cannot specify correlated observation errors for the LETKF in the KENDA system yet (refer to Sect.~\ref{sec:conclusion}). 

\begin{table*}[htb]
\caption{Observation errors and initial values for the adaptive bias correction scheme for the MeteoSwiss MWRs (assimilated channels), derived from O-B statistics. All values are brightness temperatures in \si{\kelvin}.} \label{tab:mch_mwr_settings}
    \centering 
    \begin{tabular}{l|rrrrr|rrrrr}
    \toprule
    \multicolumn{11}{c}{Observation error (instrument error of \SI{0.2}{\kelvin} is added to those values)} \\
    \midrule
    Channels & 1 & 2 & 3 & 4 & 7 & 10 & 11 & 12 & 13 & 14 \\
    Payerne  & 3.6 & 3.5 & 3.0 & 2.1 & 1.3 & 0.3 & 0.3 & 0.4 & 0.5 & 0.5 \\ 
    Grenchen & 3.6 & 3.5 & 3.0 & 2.1 & 1.3 & 0.3 & 0.3 & 0.4 & 0.5 & 0.5 \\ 
    \midrule
    \multicolumn{11}{c}{Initial values for adaptive bias correction.} \\
    \midrule
    Channels & 1 & 2 & 3 & 4 & 7 & 10 & 11 & 12 & 13 & 14 \\
    Payerne  & 1.22 & 0.44 & 0.30 & -0.29 & 0.03 & 0.85 & 0.37 & 0.51 & 0.53 & 0.60 \\
    Grenchen & 1.22 & 0.36 & 0.24 & -0.28 & 0.05 & 0.97 & 0.43 & 0.56 & 0.59 & 0.66 \\
    \bottomrule
    \end{tabular}
\end{table*}

\subsection{Impact of the MWR assimilation on the brightness temperature}

In order to assess the impact of MWR observation assimilation on the model, we recomputed a two-week period, from April 24th to May 7th 2023, of the MeteoSwiss routine data assimilation cycle without assimilation of MWR observations. This experiment can be compared directly with the operational runs, as all other settings remained unchanged. On the synoptic scale, over Switzerland, this two-week period was mainly dominated by a westerly to south-westerly flow advecting unstable air masses leading to several thunderstorms. In between frontal passages, the weather was mostly dry, with some fog formations during the nights over the Swiss Plateau. The results from this period are presented in this section.

\begin{figure*}[pbht]
\includegraphics[width=0.75\textwidth]{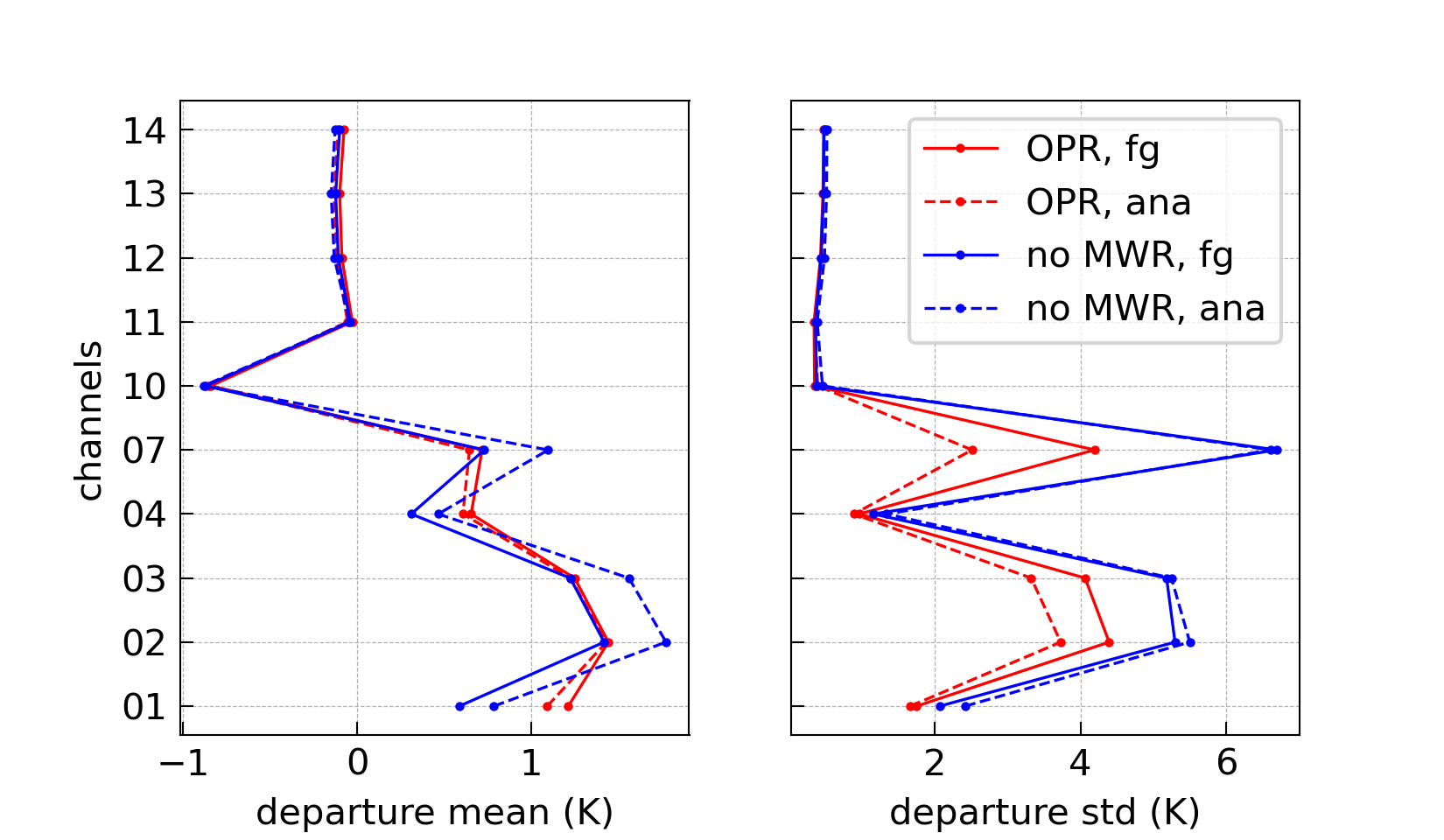}
    \caption{First guess (fg) and analysis (ana) departures in observation space for the MWR in Payerne, for the configuration with MWR assimilation (OPR; red lines) and the experiment without MWR assimilation (no MWR; blue lines). The left panel shows the mean and the right panel the standard deviation of the departures for the brightness temperatures of each assimilated channel. The statistics are sampled for all active hourly MWR measurements in the two-week period April 24th to May 7th 2023.} \label{fig:dep_tb_payerne}
\end{figure*}

First, we evaluate the performance of the MWR assimilation in observation space. Figure~\ref{fig:dep_tb_payerne} provides a summary of the mean and the standard deviation of the O-B and O-A (observation minus model analysis state) statistics aggregated over time and for different channels. When MWR observations are assimilated (operational setup), both the analysis and first guess brightness temperatures for K-band channels are closer to the observations compared to the experiment without MWR assimilation, resulting in smaller standard deviations. The largest improvement is observed in the analysis, but the impact is still visible in the first guess, indicating that the changes in the model state persist into the first forecast hour. In the setup without MWR assimilation, there is almost no difference in the brightness temperature standard deviation between the analysis and first guess, and a slight deterioration in the bias. This suggests that the state changes related to the assimilation of other observations have no positive impact on the derived brightness temperatures. 

The assimilation of MWR observations, on the other hand, demonstrates its effectiveness in influencing the brightness temperatures, indicating the successful processing of MWR observations by the LETKF. The impact on V-band channels is minimal, likely because the agreement between observations and model values is already good for those channels, limiting the potential changes introduced by the data assimilation system. The bias for V-band channels is close to zero already. However, for K-band channels, despite the adaptive bias correction scheme, a larger bias is observed. This can mainly be attributed to the fact that our cloud detection algorithm, which tends to detect clouds slightly too often, flags most of the period as being cloudy. Since the bias correction factor is not updated in cloudy conditions, it might then not fit the data well. Additionally, a two-week period, as evaluated here, may be affected by systematic model errors associated with specific weather regimes and the sample could be too small to reflect the actual bias.

The positive impact on the model first guess in terms of brightness temperature indicates that the data assimilation system is capable of changing the model state in the analysis in a way that improves the agreement between simulated brightness temperatures and observations. This encouraging result supports the direct assimilation of brightness temperature measurements.

\subsection{Impact of the MWR assimilation on the thermodynamic model state}

The evaluation in the previous section demonstrates the successful correct use of the newly introduced brightness temperature data in the KENDA system. However, the propagation of the positive impact from brightness temperature to prognostic model variables, such as temperature and humidity, is not straightforward. The mapping of the information from brightness temperature measurements to model variables highly depends on the structure of the model ensemble covariances. Therefore, it is of interest to verify the experiment and reference against other observations. For this purpose, we evaluate the quality of the COSMO-1E first guess, i.e. an hourly forecast with a one-hour lead time, for the prognostic variables of interest: temperature and humidity. For comparison, we use the \SI{00}{UTC} and \SI{12}{UTC} radiosounding observations from the Payerne station. As the first guess model states at \SI{00}{UTC} and \SI{12}{UTC} are not influenced by the radiosonde observations, the impact of the sounding assimilation on the model is 12 hours old. Hence, the sounding assimilation has no dominant impact on the verification results. 

\begin{figure*}[hbt]
\includegraphics[width=0.35\textwidth]{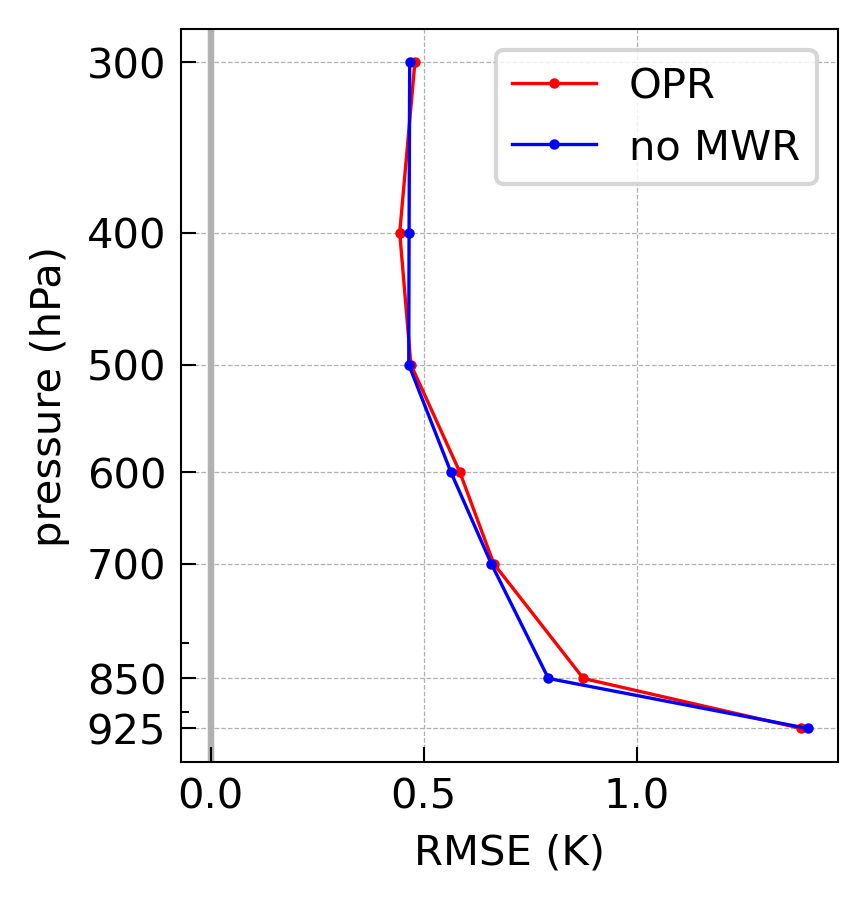}
\includegraphics[width=0.35\textwidth]{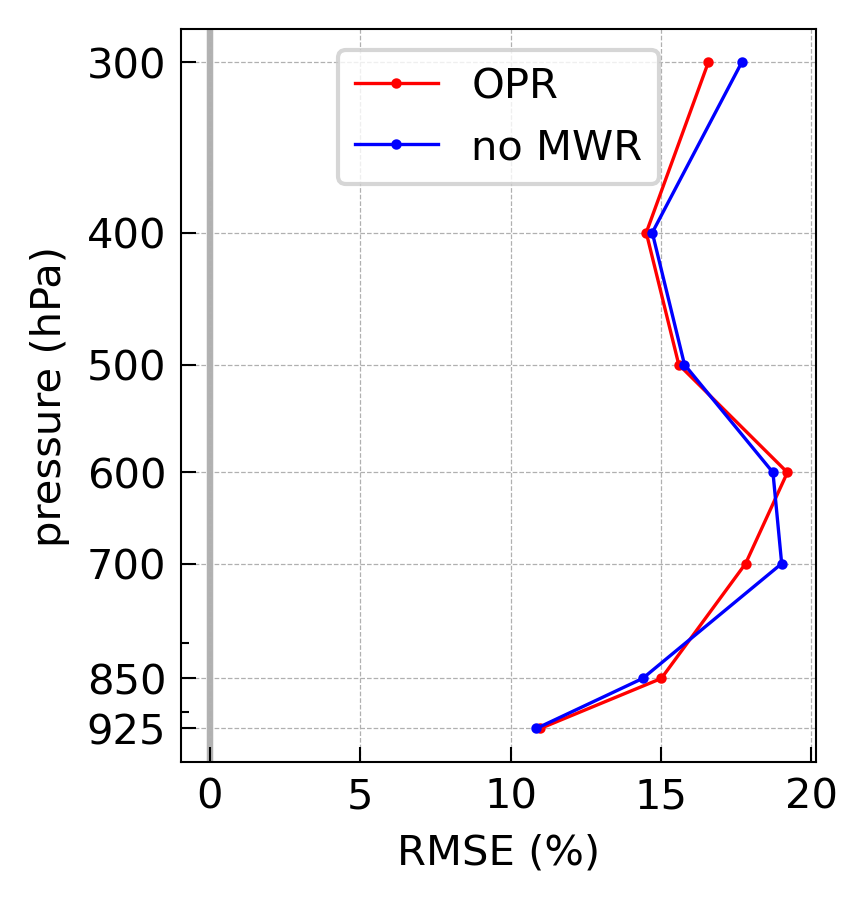}
    \caption{First guess departures for the sounding in Payerne, for the configuration with MWR assimilation (OPR) and the experiment without MWR assimilation (no MWR). The left panel shows the profile of the root mean square error of the departures for the atmospheric temperature, the right panel for the relative humidity. The statistics are sampled for all available 00 UTC and 12 UTC soundings in the two-week period April 24th to May 7th 2023.} \label{fig:dep_sounding_payerne}
\end{figure*}

Figure~\ref{fig:dep_sounding_payerne} illustrates the root mean square error (RMSE) of first guess departures for atmospheric temperature and relative humidity. Although the RMSE values differ between the two setups, the assimilation of MWR observations seems to have a neutral impact in total. To exclude the possibility that the results for relative humidity are dominated by the temperature signal, we also investigated the RMSE of specific humidity (not shown). The behaviour is the same as for relative humidity and, thus, seeming correlations between temperature and relative humidity are negligible.
Experiments we performed before the operational implementation of MWR observations in our KENDA system showed similar results when verifying forecasts with longer lead times (not shown here).

\begin{figure*}[hbt]
\includegraphics[width=0.28\textwidth]{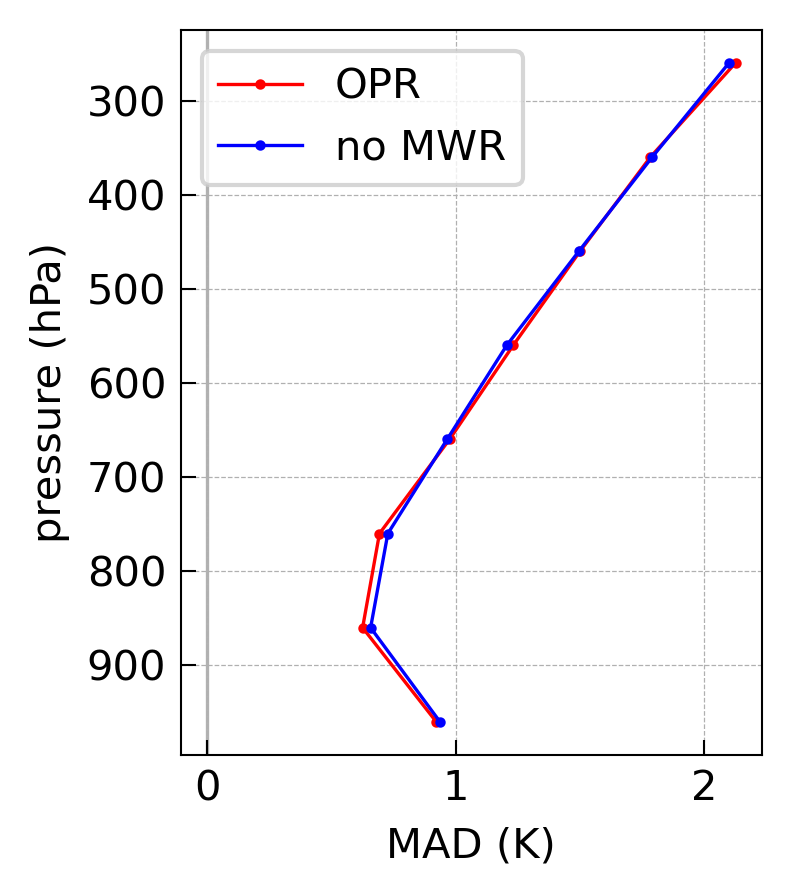}
\includegraphics[width=0.28\textwidth]{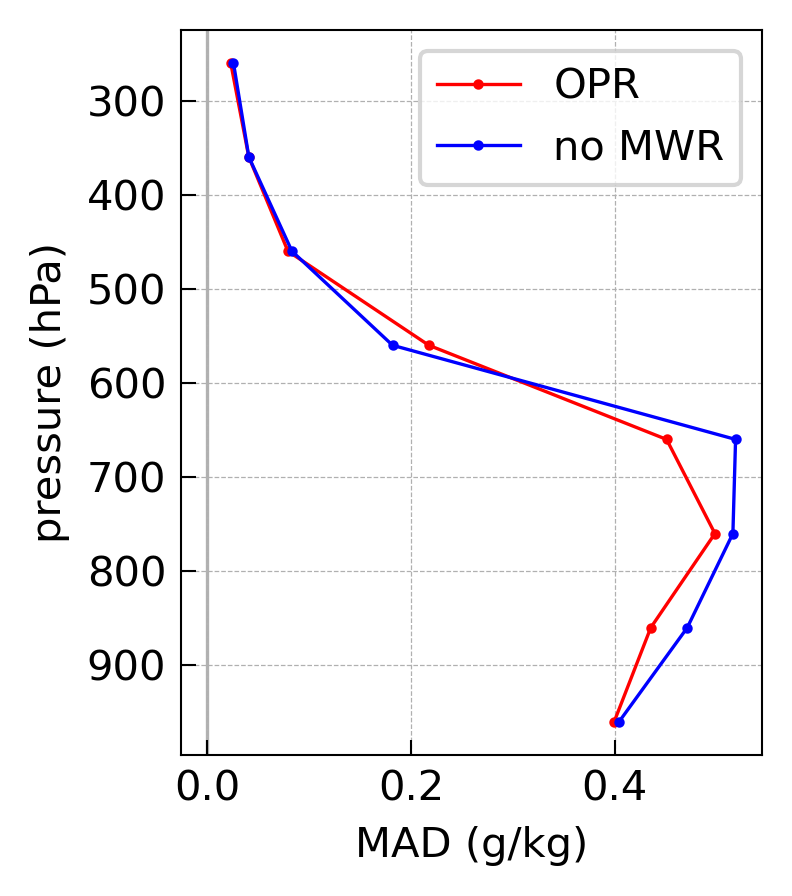}
\includegraphics[width=0.28\textwidth]{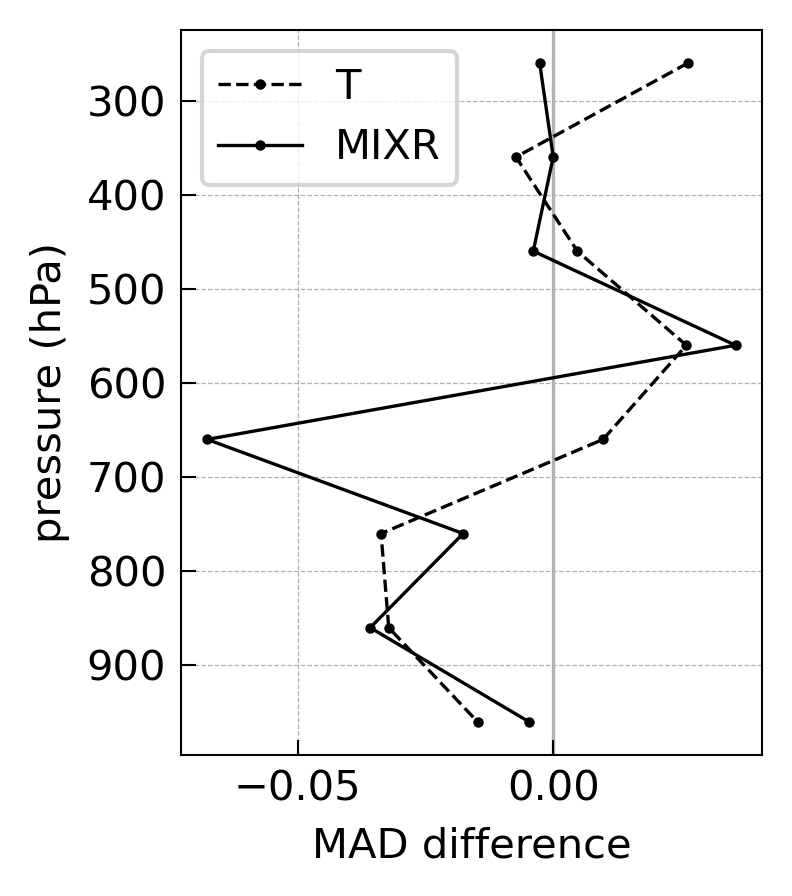}
    \caption{Departures for the Raman lidar in Payerne, for the configuration with MWR assimilation (OPR) and the configuration without MWR assimilation (no MWR). The left panel shows the profile of the median absolute deviation (MAD) of the departures for the atmospheric temperature, the panel in the middle for the mixing ratio, and the right panel the difference (OPR minus no MWR, units are \si{K} and \si{g/kg}, respectively). The statistics are sampled for all available Raman lidar data in the two-week period  April 24th to May 7th 2023.} \label{fig:dep_ralmo_payerne}
\end{figure*}

The Payerne site is equipped with a Raman lidar, which allows to extract temperature and water vapour mixing ratio profiles of the atmosphere. Currently, this data is not used for data assimilation, making it a valuable source of independent measurements. Figure~\ref{fig:dep_ralmo_payerne} presents the verification of the operational model run and the experiment without MWR observation assimilation against Raman lidar data. The difference between the operational run and the experiment is small, consistent with the verification against radiosounding. However, there is a consistent signal of improvement in the median absolute deviation (a robust measure for the variability of a sample with large outliers) in the lower part of the atmosphere for the mixing ratio. This signal indicates the potential of MWR observations to enhance the representation the ABL in the model, even though this potential is not fully utilised yet. The small impact on temperature may be attributed to the large number of already assimilated observations from MODE-S and AMDAR aircraft observations and to the assimilation of only zenith scan observations currently, while a combination of different scanning angles from the MWR contains most of the information on the temperature profile \citep{2007creloh,2009lohtur}.

MWR brightness temperature observations are complex measurements and only indirectly provide information about temperature and humidity profiles. Although the current operational assimilation of MWR observations at MeteoSwiss does not show significant improvement in model quality, it is encouraging to observe a clear improvement in brightness temperature in both the analysis and the first guess in the data assimilation cycle. Additionally, the improvement in mixing ratio when compared to Raman lidar measurements in Payerne is promising. Further investigations, as presented in the following Section~\ref{sec:exps_dwd}, will contribute to a better understanding of the assimilation process and facilitate the development of an optimised setup that fully leverages the information contained in the MWR observations in data assimilation.

\section{Assimilation experiments at DWD} \label{sec:exps_dwd}

Having obtained already a neutral to positive impact with a rather heuristic setup at MeteoSwiss, DWD pursued a different approach exploiting the strength of localisation in the LETKF but reducing the number of assimilated channels to further extract and understand the benefits of the MWR observations.

At DWD, our aim is to probe the system in a way that simulates future employment of MWRs in areas with only sparse observational coverage in the vicinity. Thus, we did not assimilate observations of the established profilers at MOL-RAO, radiosonde and radar-wind profiler (RWP), in our experiments to evaluate the benefit of potential new MWRs on sites without any profiling instrument. 

The key parameters utilised for tuning the assimilation system are the selection of assimilated channels, the observation errors, the vertical localisation, and the use of observations during cloudy conditions.
The details of the general settings of our assimilation system are described in Table~\ref{tab:das}. 
The analysis is computed once per hour and includes the following conventional observations: radar-volume data of reflectivity and radial wind, radiosonde, RWP, SYNOP, AMDAR and MODE-S aircraft observations, including the data from Berlin airport less than \SI{50}{\kilo\meter} from MOL-RAO.
Every six hours (\SIlist{00;06;12;18}{UTC}) a forecast is launched being initialised with the analysis.

For the verification, we compare the forecasted profile variables of main interest (temperature, humidity) with radiosonde observations in the close-by area of the MOL-RAO. Ascents are available every 6 hours, the descents in a similar but shifted and more variable frequency. We considered only the first \SI{6}{\hour} lead times for the verification as the impact of a single instrument on the forecast is expected to vanish after few hours. The descents have usually different coordinates due to the wind shift, thus, all observations within a radius of \SI{55}{\kilo \metre} are taken into account.

Our assimilation experiments took place in June 2021, where advection and dynamic forcing were weak and the fraction of clear-sky conditions was high at the MOL-RAO. Here, we will focus on the most important findings that led to a positive impact of the MWR observations on the forecast.

\subsection{Channel selection and observation error} \label{sec:chans} 

As described in Sect.~\ref{sec:obs}, the MWR provides measurements on 14 channels in two bands. Within each band, the observations of the single channels are partially highly correlated with each other and, thus, the observation system has few degrees of freedom only (\SIrange{1}{2}{} for humidity and \SIrange{2}{3}{} for temperature) when not including elevation scans \citep{2009lohtur}. Similarly, it can be assumed that the observation errors $\epsilon_\mathrm{o}$ exhibit interchannel cross-correlations, which also includes errors introduced by the forward operator. In principle, the LETKF can deal with cross-correlated observation errors by taking into account also the off-diagonals of the observation-error covariance matrix ($\pmb R$). The KENDA system, however, can currently only treat the observation-error variances -- the treatment of the covariances is not implemented into the system yet.

A consequence of neglected interchannel-cross correlations can be an underestimation of the observation errors when assimilating many channels simultaneously. 
Accordingly, one of the most important aspects that improved the impact of the MWR observations on the forecast was a reduction of assimilated channels. We performed experiments with 14, 11, 6, 4, and 2 channels and found the two-channel setup to exhibit the best results. In the following, we will present only results that include one channel from each of the two bands. We chose channel 2 and 13 (ch2, ch13) since they are located in different spectral regimes and show good O-B statistics (cf. Sect.~\ref{sec:omb}), while still being representative for the sensitivity characteristics of their respective spectral band. 

The two-channel setup also allowed for a more reliable computation of the
observation errors $\epsilon_\mathrm{o}$ using the Desroziers diagnostics \citep{2005desber}. After several iterations, we found that $\epsilon_\mathrm{o,ch2}$ \SI{{\approx} 1.0}{\kelvin} and $\epsilon_\mathrm{o,ch13}$ \SI{{\approx} 0.25}{\kelvin}.
As the correlations between the observation errors of those two channels are negligible, the diagnosed error values do not change significantly for presented single-channel experiments.

\begin{table*}[htb]
\caption{Assimilation experiments discussed in Sects.~\ref{sec:vloc} and \ref{sec:scexp}. Vertical localisation height $p_\mathrm{L}$ and width $dp_\mathrm{L}$ for each channel. Assimilation for all-sky is indicated by yes or no, respectively. The values for KV1 indicate the temporally averaged values of the dynamical localisation based on the sensitivities of the forward model.} \label{tab:exp}
    \centering \small
    \begin{tabular}{l|rrl|rrl}
    \toprule
    name & $p_\mathrm{L}$ (\si{\hecto \pascal}) & $dp_\mathrm{L}$ & all-sky & $p_\mathrm{L}$ (\si{\hecto \pascal}) & $dp_\mathrm{L}$ & all-sky \\ 
             & \multicolumn{3}{c|}{channel 2} & \multicolumn{3}{c}{channel 13} \\ \midrule
    KV1  & $\approx$ 830 & $\approx$ 0.18 & no & $\approx$ 970 & $\approx$ 0.075 & no  \\ 
    KV2  &           760 &           0.15 & no &          1000 &            0.15 & no  \\ 
    KV3  &           590 &           0.15 & no &          1000 &           0.075 & no  \\ 
    K1   &           590 &           0.15 & no &             - &               - &   - \\ 
    K2   &           760 &           0.12 & no &             - &               - &   - \\ 
    K3   &           590 &            10  & no &             - &               - &   - \\ 
    V1   &             - &              - &  - &          1000 &           0.075 & yes \\ 
    V2   &             - &              - &  - &          1000 &            0.15 & yes \\ 
    V3   &             - &              - &  - &           900 &              20 & yes \\ 
    \bottomrule
    \end{tabular}
\end{table*}

\subsection{Vertical localisation} \label{sec:vloc}

\begin{figure*}[htb]
\includegraphics[height=0.27\textheight]{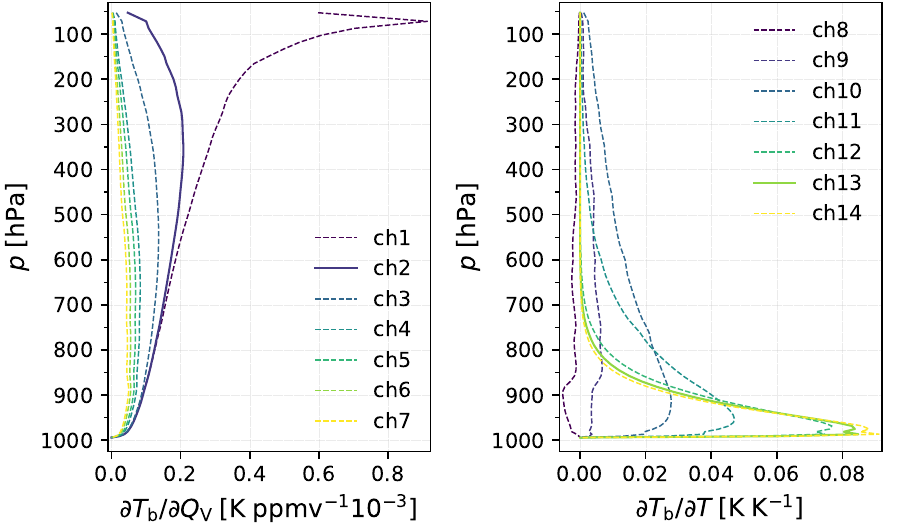}
\includegraphics[height=0.27\textheight]{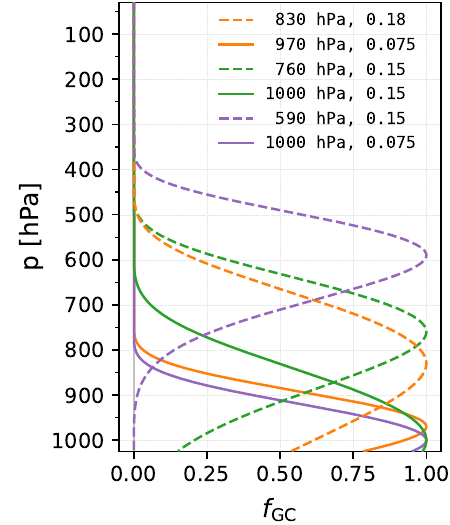}
    \caption{Jacobians computed with RTTOV-gb for sensitivities of K-band channels to humidity (left) and of the V-band channels to temperature (centre). The channels studied here in more detail (ch2, ch13) are indicated by a solid line. Sensitivities of K-band channels to temperature and V-band channels to humidity are rather small and therefore not shown here for clarity. Right: Gaspari-Cohn functions for ch2 (dashed) and ch13 (solid) representing the vertical localisations employed in the two-channel experiments (cf. Table~\ref{tab:exp}), namely KV1 (orange), KV2 (green), and KV3 (purple).} \label{fig:gcf}
\end{figure*}

Localisation can be central to ensemble data assimilation as it not only reduces spurious correlations but also increases the effective size of the ensemble space \citep{2001houmit} compared to the assimilation of unlocalised observations. Vertical localisation of profile-integrated measurements is, however, not straightforward and still subject to research.
Non-local observations like passive MWR measurements are posing a problem for the LETKF where localisation is performed in observation space by reducing the weight given to an observation in the assimilation process according to a Gaspari-Cohn-function \citep{1999gascoh} $\pmb f_{\mathrm{GC}}(p_{L},dp_{L})$. This method particularly requires a localisation pressure level $p_{L}$ and localisation width $dp_{L}$ being assigned to each assimilated observation.

In our first experiments, we computed $p_{L}$ and $dp_{L}$ dynamically using weighting functions \citep{2020hutsch} that are based on the sensitivity functions (Jacobians) of RTTOV-gb (cf. \citealt{2016decim}). Channel 13 shows a clear sensitivity maximum close to the ground (Fig.~\ref{fig:gcf}, centre), the derived $p_\mathrm{L}$ are around \SI{970}{\hecto \pascal} (Fig.~\ref{fig:gcf}, right). While the sensitivity of the water vapour channels is distributed quite broadly over the troposphere (Fig.~\ref{fig:gcf}, left), the corresponding localisation levels $p_{L}$ are still at relatively low altitude (around \SIrange{810}{850}{\hecto \pascal} for ch2; Fig.~\ref{fig:gcf}, right) as its computation involves the product of the Jacobian with the vertical distribution of the variable itself.

\begin{figure}[hbt]
    \includegraphics[width=0.5\textwidth]{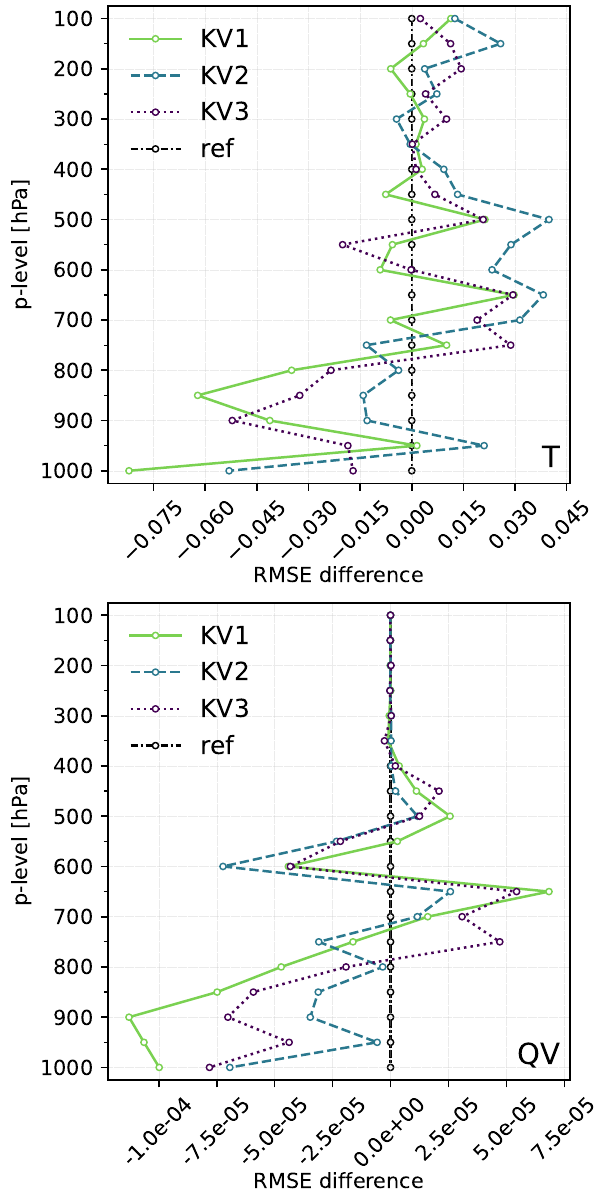}
    \caption{Verification of the forecast against radiosonde measurements. RMSE difference for T (top) and QV (bottom) between the respective experiments and the reference run summarised over all lead times until \SI{6}{\hour} for all two-channel experiments.}
    \label{fig:verisum1}
\end{figure}

Adopting this dynamic vertical localisation for ch2 and ch13 (Table~\ref{tab:exp}) yields a positive impact in the forecast verification of humidity in the ABL while the impact on the temperature is more mixed but also mostly positive in the ABL (Fig.~\ref{fig:verisum1}; KV1).

Some assessment of the adequacy of the chosen localisation heights can be obtained from the cross-validation diagnostics presented by \citet{2022sti}. The most prominent of these diagnostics indicates whether the assimilation of an observation (e.g., by the MWR) would pull the analysis towards other well-trusted (``verification-'') observations. The way we employed this method here amounts to an assessment of the impact of the MWR assimilation on the analysis with radiosonde observations as verification data. More specifically, for computational simplicity, we are using the \textsl{single observation version} of the diagnostics from \citet{2022sti} which yields the impact that would be obtained if the MWR were assimilated in single observation experiments, respectively.

\begin{figure}[hbt]
    \includegraphics[width=0.24\textwidth]{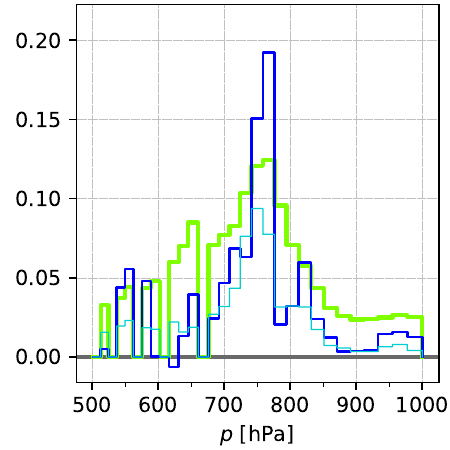}
    \includegraphics[width=0.24\textwidth]{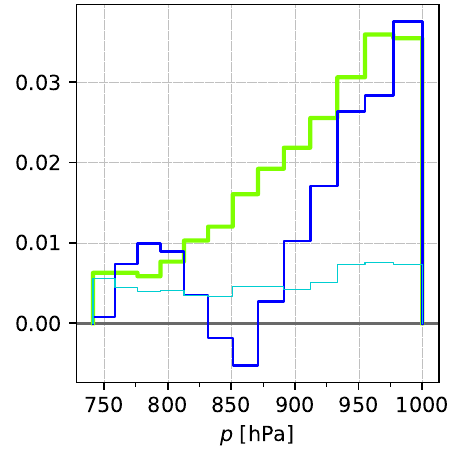}
    \caption{Cross validation diagnostics for the KV1 experiment (cf. Table~\ref{tab:exp}). Diagnostics related to observational (blue) and employed (green) covariance estimates against pressure level (see Sect.~\ref{sec:vloc} for details).} The distributions are cropped on the left side of the x-axis as the level of noise (cyan) becomes too large here. Left: ch2 vs. relative humidity. Right: ch13 vs. temperature.  \label{fig:crossv}
\end{figure}

The blue curves in Fig.~\ref{fig:crossv} show the actual impact as assessed by the observations while the green curves indicate the corresponding impact which would be obtained if the error covariances used (implicitly or explicitly) in the data assimilation system were fully correct (and if biases could be neglected). Here, positive values indicate a beneficial impact (i.e., consistency between MWR assimilation and the radiosonde observations). As indicated on the x-axis, statistics are collected in bins corresponding to the pressure level of the respective radiosondes. 

Figure~\ref{fig:crossv} presents the diagnostics applied to the experiment where $p_{L}$ and $dp_{L}$ were computed from the Jacobians. While for the temperature-sensitive ch13 these data have their maxima in the bin with the lowest levels and therefore suggest a localisation level close to the ground (which is in agreement with the levels computed from the Jacobians), the humidity-sensitive ch2 is shown to have the by far most positive impact around \SI{760}{\hecto \pascal} while the values of $p_{L}$ in this experiment were mainly between \SIrange{810}{850}{\hecto \pascal}. 

These diagnostics motivated experiment KV2, where the derived estimates for $p_{L}$ and $dp_{L}$ were adopted as static values for the vertical localisation of the channels. 
This experiment, however, exhibited much higher forecast errors. In particular, the RMSE improvements seen in the lower-level temperature and humidity of KV1 have been strongly reduced in KV2 (Fig.~\ref{fig:verisum1}). 

This shows that a straight forward application of the cross-validation diagnostics for deriving localisation parameters can be problematic which may have various reasons. 
It has to be stressed that the applied cross-validation diagnostics show only the impact on the analysis and that it is not trivial to assess how different aspects of the analysis affect the forecast. Furthermore, the small amount of data related to the single MWR instrument available in this study requires some caution when interpreting details. An indicator for the statistical significance of the cross-validation diagnostics (blue curves) is given by the cyan curves in Fig.~\ref{fig:crossv}, which yield the variance of a random process with zero mean whose increments have the same size as for the corresponding blue curve but for which the sign (+ or -) of the increments is completely random. Particularly in the left graph (for ch2) the blue curve is quite close to this noise indicator (rarely exceeding the cyan curve by more than a factor 2). While the fact that the blue curve has positive values in almost all bins still suggests that the overall impact on the analysis is beneficial, the proximity of the blue and cyan curves shows that the values in the individual bins have to be interpreted with great caution. We hope that in future studies with a larger amount of available data, such diagnostics will give better guidance with greater statistical significance.

For testing the influence of cross-correlations between ch2 and ch13, we configure the localisation of a third experiment (KV3) in a way that the localisation regions basically have no vertical overlap.
Compared to KV2, most of the positive impact from KV1 seems to be regained apart from some degradations for single levels, especially for the bottom-most temperature level.

\subsection{Single channel experiments}\label{sec:scexp}

In the previous section, we presented experiments, where we assimilate two channels simultaneously. We have obtained a positive impact by using the approach of deriving the localisation functions from the model sensitivities (KV1). The other employed approaches (KV2 and KV3) were not able to further improve these results. 
To better understand the forecast impact from each of the two channels and to further investigate the influence of the respective vertical localisations, we performed a couple of single-channel experiments with different localisation settings (Table~\ref{tab:exp}) for both ch2 and ch13. 

As the O-B statistics shown in Sect.~\ref{sec:omb} indicated a low sensitivity of ch13 to cloudy conditions, we assimilate this channel for all-sky and increase the observation error by a factor of 1.4 in the presence of clouds according to the STD ratio derived from the respective cloudy-/clear-sky O-B. This led in general to a slight increase of the positive impact for the ch13-only experiments compared to clear-sky experiments (not shown here). The average improvement in the lower part of the atmosphere ($<$\SI{650}{\hecto\pascal}) is 1.9\%/5.1\%/2.2\% for the temperature RMSE of each of V1/V2/V3 compared to the clear-sky experiment. For humidity, V1 is degraded by 0.60\% and V2/V3 improved by 2.3\%/4.1\% on average.

\begin{figure*}[phtb]
    \includegraphics[width=\textwidth]{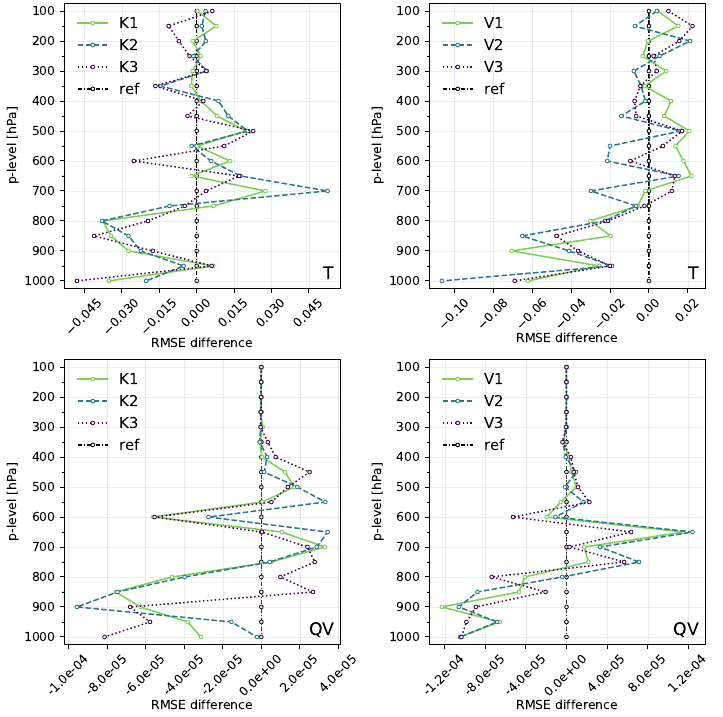}
    \caption{Verification of the forecast against radiosonde measurements. RMSE difference for T (top) and QV (bottom) between the respective experiments and the reference run summarised over all lead times until 6h for all described ch2 (left) and ch13 (right) experiments.} \label{fig:veri}
\end{figure*}

All ch2-experiments (Fig.~\ref{fig:veri}, left) show a similar impact on the temperature. Most strikingly, we do not observe significant differences between experiments with unlocalised and sharp vertical localisation. 
The largest differences for humidity are visible in the lower ABL. Here, the unlocalised K3 shows the smallest RMSE near the surface but degrades rapidly for higher levels. 

All ch13-experiments improve the RMSE below \SI{800}{\hecto \pascal} for both temperature and humidity (Fig.~\ref{fig:veri}, right). The RMSE of humidity around \SI{700}{\hecto \pascal} seems to be difficult to improve as none of the experiments shows a positive score in that area. We suspect that this could be a consequence of the good performance of the reference at these levels. 
%
To further assess the role of the performance of the reference and, thus, the reliability of the results, we performed a consistency check against another reference with only slightly different settings (marginally smaller domain and RWP assimilation switched on at MOL-RAO; not shown here). As both references should differ mostly at the level of statistical noise, we would expect the same behaviour of our experiments in comparison.

And indeed, we see a consistent positive impact for only a part of the presented verification data, e.g., for the lowest temperature level for all ch2-experiments. The impact of K3 on the lowest humidity level seems consistent as well as levels around \SIrange{900}{800}{\hecto\pascal} for K1 and K2. This differentiated view on the impact indicates that the results for this set of experiments have to be taken with care.
Due to the observed change of sign of the RMSE difference, the mentioned seemingly negative impact of the ch13-experiments on humidity around \SI{700}{\hecto \pascal} seems to be more of an effect of statistical noise as well. In contrast, the ch13-experiments show a consistent positive impact for the lowest levels for both temperature and humidity. These more trustworthy results could be an effect of the all-sky assimilation for ch13, which almost doubles the number of actively assimilated observations in the investigated period. 
In contrast to the findings of Sect.~\ref{sec:vloc}, the exact vertical localisation in the presented single-channel experiments seems to be only a second-order effect though.

\begin{figure}[htb]
    \includegraphics[width=0.5\textwidth]{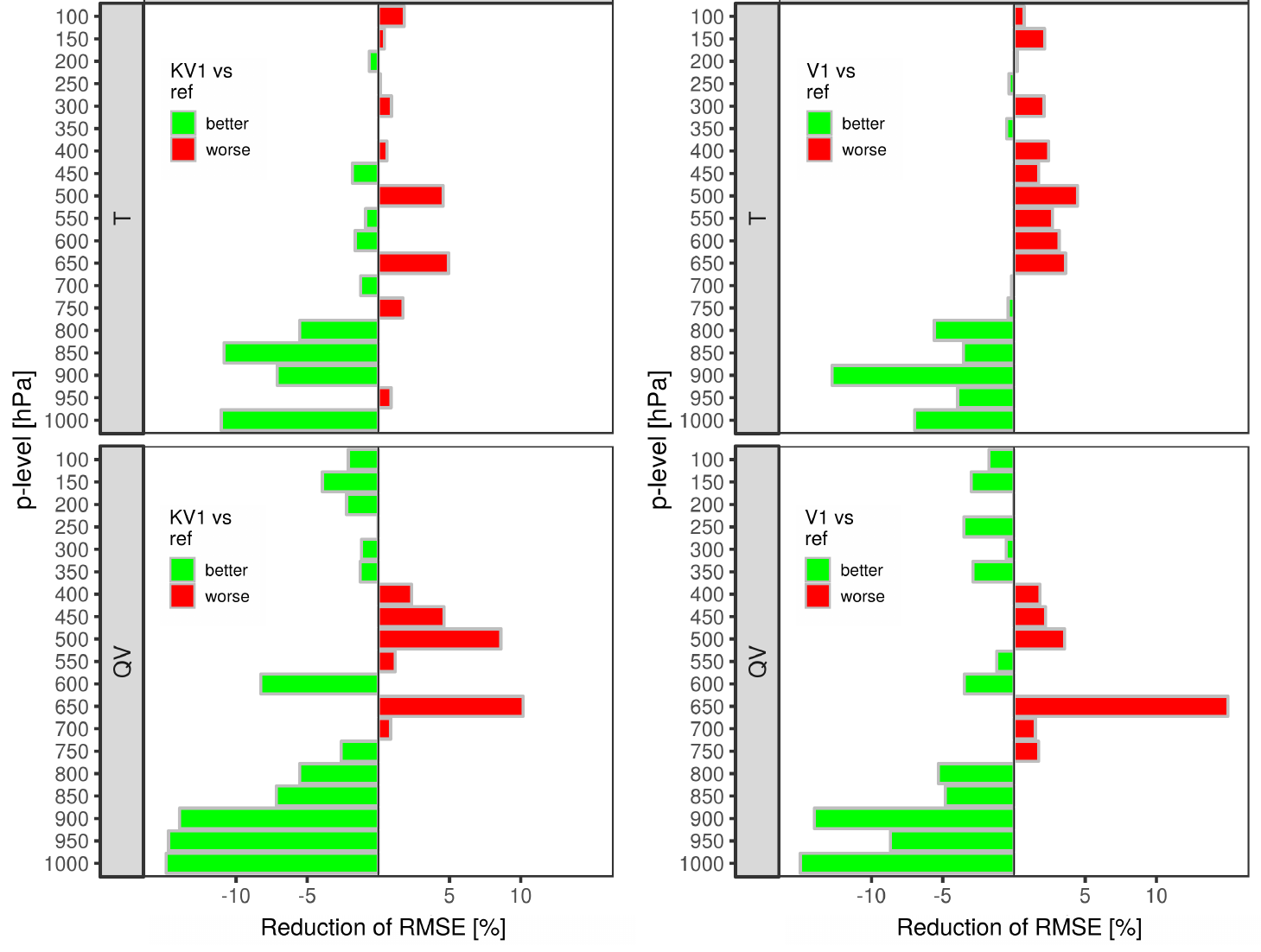}
    \caption{RMSE improvement (in \%) for T (top) and QV (bottom) for KV1 (left) and V1 (right), respectively, vs. the reference run summarised over all lead times until \SI{6}{\hour}.}
    \label{fig:verisum}
\end{figure}

When comparing one of the ch13-experiments with the best two-channel experiment KV1, they perform very similarly within the expected noise range, with both having a positive impact in most of the ABL. 
Seemingly larger differences between the RMSE summaries of KV1 and, e.g., V1 (Fig.~\ref{fig:verisum}) have only limited significance.
Apparently, the larger amount of information contained in the temperature-sensitive channel when assimilating for all-sky can make up for the lack of the more direct humidity information provided by ch2.
Additionally performed two-channel all-sky experiments with the already presented localisation settings were not able to further improve the forecast verification consistently though.

We would like to stress again that our results have to be taken with care due to the already mentioned small number of verifying measurements and the fact that we evaluate only the impact of observations of a single instrument here (cf. Sect.~\ref{sec:vloc}). Differences between the experiments generally have a low significance due to the temporally sparse verifying radiosonde observations. The bulk of radiosonde observations is available at lead times of \SI{5}{\hour}, where the largest impact of the MWR observations might have vanished already. Descents take place mostly at \SI{2}{\hour} already but are scattered in the area around the MOL-RAO.

\section{Conclusion} \label{sec:conclusion}

We presented our progress on the use of MWR observations in the assimilation systems of MeteoSwiss and DWD. We implemented the direct assimilation of the brightness temperatures employing RTTOV-gb as forward operator and investigated the impact of the updated initial conditions on the NWP forecasts.

MWR observations have been assimilated in the operational setup at MeteoSwiss since the summer of 2022. Although the impact on the forecast quality of the meteorological variables is mostly neutral, we have noticed a slight improvement in the lower ABL when comparing the first guess to observations of an independent Raman lidar. However, it is important to note that the current setup does not utilise all available MWR channels and elevation scans. Furthermore, we did not apply vertical localisation to the assimilated MWR observations. This localisation can have a non-negligible impact on assimilation results, as discussed in Section \ref{sec:vloc}. Therefore, we believe that there is still more potential for assimilating MWR data, and further adjustments to the system could lead to improvements. After the migration from the COSMO to the ICON model, a reassessment of the MWR data assimilation settings can take place in considerations of the results found at DWD.

In the presented DWD assimilation experiments, we focused on the investigation of the impact of the vertical localisation of each of the two selected channels, the humidity-sensitive ch2 and the temperature-sensitive ch13. 
We showed that a proper vertical localisation can be important when assimilating both channels simultaneously but has a secondary effect on the RMSE of the forecast verification when assimilating only single channels.
In our experiments, actually, ch13 alone is sufficient to yield a good impact on the forecast of both temperature and humidity in the ABL. These experiments yield approximately the same impact as the two-channel experiment using the dynamic localisation (KV1), which could be an indication that part of the clear-sky humidity information of ch2 can be compensated by the larger amount of assimilated ch13 observations due to the all-sky assimilation.

Our results show that the direct assimilation of MWR brightness temperatures is indeed able to improve the NWP forecast, especially in the ABL. Nevertheless, due to the assimilation of only one instrument and the lack of more frequent verifying observations, our results are often difficult to interpret in more detail.

In addition, the integration of more channels, which could provide more information on the profiles, is challenging due to the more complex height assignment and the fact that accounting for observation-error cross-correlations is not yet available in the assimilation system. 
Further improvement of the observation impact, especially by the use of more channels, will most likely be possible with the implementation of the full $\pmb R$-matrix into the KENDA system, which will allow for a treatment of the correlated observation errors. Developments in this direction are ongoing at DWD.

These developments are also important for potential future assimilation experiments, where one might also assimilate the non-zenith angles of the temperature-sensitive channels. This would add more degrees of freedom to the system on the one hand, but, on the other hand, additional correlations between observations of different angles have to be expected. Moreover, the detection and treatment of cloudy conditions is more complex for non-zenith angles as the line of sight might cross several model grid cells and detected clouds might be located outside of the considered grid cell.
Since RTTOV-gb assumes a horizontally homogeneous atmosphere to compute the slanted scans, a detailed analysis of the deviations between observations and model equivalents in those cases is needed.

As already mentioned, the evaluation of the impact of observations of single instruments on the forecast suffers generally from the difficulty to produce significant results. Future experiments might be able to take into account more MWR observations in the growing European network fostered by, e.g., E-PROFILE and ACTRIS.

At MeteoSwiss, a third MWR station, close to Schaffhausen, will be implemented into the system after an upgrade of the instrument in 2023. DWD started to operate a second MWR at the weather station in Aachen-Orsbach to enable an end-to-end testing in an operational framework. MOL-RAO is currently evaluating a compact differential absorption lidar (DIAL) and a Raman lidar. These profilers can also help addressing the need for a more frequent reference for the verification of humidity and temperature forecasts compared to radiosoundings.

\section*{Acknowledgements}
The authors would like to thank Bas Crezee for his valuable support in conducting the comparison between the MeteoSwiss model runs and the Raman lidar observations, and Giovanni Martucci for providing quality controlled observations for this purpose. We are also grateful to Hendrik Reich for his extensive support on various technical aspects of the assimilation system.

Jasmin Vural was funded by the Deutsche Forschungsgemeinschaft (DFG, German Research Foundation) -- 399851006.

Numerical simulations conducted by MeteoSwiss are all performed at the Swiss National Supercomputing Center (CSCS).

\section*{Data Availability Statement}
The data that support the findings of this study are available from the corresponding author upon reasonable request.

\printendnotes

\small
\bibliography{bibexport} 

\end{document}